\documentclass[preprint,floatfix,nofootinbib,superscriptaddress,12pt]{elsarticle}

\usepackage{graphicx,subfig,epsfig,epsf,color}
\usepackage{amssymb,amsmath}
\usepackage{slashed}
\usepackage{array}
\usepackage{hyperref}

\journal{Nuclear Physics A}

\begin{document}

\begin{frontmatter}



\title{Constraints on Density Dependent MIT Bag Model Parameters for Quark and Hybrid Stars}

\author[vecc,hbni]{Soumen Podder}
\ead{s.podder@vecc.gov.in}

\author[vecc,hbni]{Suman Pal}
\ead{sumanvecc@gmail.com}

\author[ku]{Debashree Sen\corref{cor1}}
\ead{debashreesen88@gmail.com}

\author[vecc,hbni]{Gargi Chaudhuri}
\ead{gargi@vecc.gov.in}

\cortext[cor1]{Corresponding author}

\address[vecc]{Physics Group, Variable Energy Cyclotron Centre, 1/AF Bidhan Nagar, Kolkata 700064, India}

\address[hbni]{Homi Bhabha National Institute, Training School Complex, Anushakti Nagar, Mumbai 400085, India}

\address[ku]{Center for Extreme Nuclear Matters (CENuM), Korea University, Seoul 02841, Korea}

\begin{abstract}

We compute the equation of state (EoS) of strange quark stars (SQSs) with the MIT Bag model using density dependent bag pressure, characterized by a Gaussian distribution function. The bag pressure's density dependence is controlled by three key parameters namely the asymptotic value ($B_{as}$), $\Delta B(=B_0 - B_{as})$, and $\beta$. We explore various parameter combinations ($B_{as}$, $\Delta B$, $\beta$) that adhere to the Bodmer-Witten conjecture, a criterion for the stability of SQSs. Our primary aim is to analyze the effects of these parameter variations on the structural properties of SQSs. However we find that none of the combinations can satisfy the NICER data for PSR J0030+0451 and the constraint on tidal deformability from GW170817. So it can be emphasized that this model cannot describe reasonable SQS configurations. We also extend our work to calculate structural properties of hybrid stars (HSs). With the density dependent bag model (DDBM), these astrophysical constraints are fulfilled by the HSs configurations within a very restricted range of the three parameters. The present work is the first to constrain the parameters of DDBM for both SQS and HSs using the recent astrophysical constraints on tidal deformabiity from GW170817 and that on mass-radius relationship from NICER data.

\end{abstract}





\end{frontmatter}


\section{Introduction}
\label{Intro}

The physics of dense matter related to compact stars (density $\rho$=5-10 times the nuclear density) is quite inconclusive specially in terms of composition and nature of interactions. Therefore, whether quark matter (QM) can be a possible candidate at such dense environment is still an open question and one of the current topics of interest. Theoretically, two possible scenarios can possibly support the concerned topic - i) phase transition from hadronic to QM forming hybrid stars (HSs) \cite{Glendenning:1997wn,Blaschke:2020vuy,Ferreira:2020evu,Zha:2020gjw,Lugones:2021tee,Husain:2020nbb,Morimoto:2020fkg,Clevinger:2022xzl,Pal:2023quk,Sen:2022lig,Sen:2021cgl,Sen:2022qol} and ii) QM may be the absolute ground state of matter that interacts strongly at such density. Strange Quark Matter (SQM), being composed of u, d and s quarks, being more stable than pure nucleonic system, is the true ground of the matter (Bodmer-Witten conjecture) \cite{Farhi:1984qu,Torres:2012xv} which  implies that the entire compact star may be composed of SQM and thereby  leading to formation  of strange quark stars (SQSs). Consequently, several theoretical models came up with the attempt to describe the possible existence of SQM in SQSs. The first one was the original MIT Bag model \cite{Chodos:1974je}. Later repulsive interaction between quarks was introduced via a parameter $\alpha_4$ \cite{Weissenborn:2011qu} and also by introducing vector meson as mediator (vBag model) \cite{Cierniak:2019hhe,Lopes:2020btp,Lopes:2020dvs,Kumar:2022byc}. Gluon effects were also introduced in the MIT bag model \cite{Albino:2021zml}. Attempts were made to understand the properties of QSs with such original and modified MIT bag model \cite{Miao:2021nuq,Li:2020wbw}. Several other models like the quark-mass density dependent model \cite{Lugones:2022upj}, the Nambu-Jona-Lasinio model \cite{Li:2019ztm}, the quasi-particle model \cite{Li:2019akk}, the Dyson-Schwinger model \cite{Luo:2019dpm}, the Polyakov Chiral SU(3) quark mean field model \cite{Kumari:2021tik}, interacting QM that includes inter-quark effects from perturbative QCD and color superconductivity \cite{Zhang:2020jmb} etc. have also been adopted for the same purpose.
 
 In the present work we consider the simplest form of the MIT Bag model without involving the perturbative or repulsive effects to compute the equation of state (EoS) and structural properties of SQSs as well as HSs. The MIT Bag model is  characterized by a bag pressure $B$, which is often considered to be constant and independent of density. However, it is already known that the original/simplest form of Bag model with constant bag pressure cannot satisfy the observational constraints from massive pulsars in case of QSs \cite{Lopes:2020btp} unless repulsive or perturbative corrections are included. Moreover, at high densities the quarks gain asymptotic freedom. Therefore in the context of QSs where the density is very high, it is more justified to consider the bag pressure to be density dependent. In order to invoke this concept of asymptotic freedom of quarks at high density, density dependence of the bag pressure $B(\rho)$ was considered by \cite{Burgio:2001mk,Burgio:2002sn} in the form of a Gaussian distribution involving the parameters like asymptotic value ($B_{as}$) of $B(\rho)$, $B_0(\rho=0)$ and $\beta$. In \cite{Burgio:2002sn} these parameters were fixed in such a way that they generate the hadron-quark transition energy density consistent with that predicted by CERN-SPS. Clearly, the values of $B_{as}$, $B_0$ and $\beta$ considered in \cite{Burgio:2002sn} in the context of hadron-quark phase transition, are not suitable for explaining the EoS and structure of SQSs. Prior to the NICER and gravitational wave detection era, few works \cite{Burgio:2001mk,Burgio:2002sn,Bordbar:2020fqj,Bordbar:2013uwa,Miyatsu:2015kwa,Isayev:2015jqa,Yazdizadeh:2013cxa,Prasad:2003bw} considered this Gaussian form of the density dependent bag pressure to obtain SQS and/or HS properties. However, with these recent astrophysical constraints, it is very important to test the parameters of the density dependent Bag model in the context of formation of SQS and HSs. Thus considering the present literature, this is the first study of constraining the above mentioned parameters in density dependent Bag model for both SQS as well as HSs using the recent astrophysical constraints on tidal deformability from GW170817 and that on mass-radius relationship from NICER data. In the first part of the present work, we study the properties of SQSs in the light of the various recent astrophysical constraints where the value of $B_0$ also affects significantly the present results. 
 
 We also extend our work in order to investigate the role of the three parameters $B_{as}$, $B_0$ and $\beta$ in determining the structural properties of HSs with respect to the same astrophysical constraints. We consider the NL3$\omega\rho$6 model \cite{Fortin:2021umb,Grill:2014aea,Pais:2016xiu} based on the relativistic mean field (RMF) formalism for the hadronic phase. The RMF treatment has been widely adopted to construct dense matter EoS with several other chiral RMF models \cite{Li:2019akk,Li:2019ztm,Lopes:2020dvs,Morimoto:2020fkg,Kumari:2021tik,Wang:2001jw,Singh:2018kwq,Wang:2002aq}. Like SQSs, we compute the HS properties considering the bag pressure to be dependent on density. Phase transition in case of HSs is achieved with Maxwell construction assuming that the surface tension at hadron-quark interface is large enough to make the existence of mixed phase (with Gibbs construction) unstable \cite{Maruyama:2007ss}. We intend to show how far the values of $B_{as}$, $B_0$ and $\beta$ differ in the two scenarios of SQSs and HSs in order to satisfy the present astrophysical constraints. Ref. \cite{Fonseca:2021wxt} measured quite accurately the mass of the most massive pulsar PSR J0740+6620 to be 2.08 $\pm$ 0.07 $M_{\odot}$. Recently, the NICER experiment for this pulsar also found its radius along with the mass as $M$ = 2.08 $\pm$ 0.07 $M_{\odot}$ and $R = 13.7^{+2.6}_{-1.5}$ km \cite{Miller:2021qha} and $M = 2.072^{+0.067}_{-0.066} M_{\odot}$; $R = 12.39^{+1.30}_{-0.98}$ km \cite{Riley:2021pdl}. The NICER experiment also determined the mass and radius of the pulsar PSR J0030+0451 as $M = 1.34^{+0.15}_{-0.16} M_{\odot}$; $R = 12.71^{+1.14}_{-1.19}$ km \cite{Riley:2019yda} and $M = 1.44^{+0.15}_{-0.14} M_{\odot}$; $R = 13.02^{+1.24}_{-1.06}$ km \cite{Miller:2019cac}. Moreover, we also compare our results of the tidal deformability of 1.4 $M_{\odot}$ NS with that obtained from GW170817 observational data (70$\leq \Lambda_{1.4} \leq$580) utilizing the binary Love relation \cite{LIGOScientific:2018cki}. Empirically, with Gaussian process it was found $\Lambda_{1.4}=211^{+312}_{-137}$ \cite{Essick:2019ldf} at 90\% confidence level. Other empirical techniques also constrained the binary tidal deformability as $\tilde{\Lambda}$ \cite{LIGOScientific:2018hze,De:2018uhw,Essick:2019ldf} at 90\% confidence level. One can refer to \cite{MUSES:2023hyz} for the updated values of all the astrophysical constraints. In this context it is worth mentioning that in the existing literature, the variation of these parameters $B_{as}$, $B_0$ and $\beta$, used to obtain the density dependence of the bag pressure, were not tested before in terms of the constraint on $\Lambda_{1.4}$ from GW170817.

The paper is organized as follows. In the next section \ref{Formalism}, we address the MIT Bag model with density dependence of the bag pressure, mechanism of phase transition and the structural properties of SQSs and HSs. We then present our results and relevant discussions in section \ref{Results}. We summarize and conclude in the final section \ref{Conclusion} of the paper.


\section{Formalism}
\label{Formalism}

\subsection{Strange Quark Stars and MIT Bag model}
\label{Quark Phase}

We consider the MIT bag model \cite{Chodos:1974je,Glendenning:1997wn} with u, d and s quarks along with the electrons. The u and d quarks have negligible mass compared to that of the s quark ($m_s \approx$ 93.4~MeV) \cite{ParticleDataGroup:2018ovx}. We follow \cite{Glendenning:1997wn} for the general formalism for obtaining the EoS of SQM using the MIT Bag model. This model is characterized by the bag pressure $B$ is actually the energy density difference between the perturbative vacuum and the true vacuum \cite{Burgio:2001mk,Burgio:2002sn} and is often considered to be independent of density (constant) in literature. In this model, the energy density and pressure of the quarks can be expressed as \cite{Glendenning:1997wn}

\begin{eqnarray}
\varepsilon = B(\rho)+ \sum_f \frac{3}{4\pi^2} \Biggl[\mu_fk_f\Big(\mu_f^2-\frac{1}{2}m_f^2\Big) - \frac{1}{2}m_f^4 \ln\Big(\frac{\mu_f+k_f}{m_f}\Big)\Biggr]
\protect\label{eos_e_uqm}
\end{eqnarray}

and

\begin{eqnarray}
P = -B(\rho)+ \sum_f \frac{1}{4\pi^2} \Biggl[\mu_fk_f\Big(\mu_f^2-\frac{5}{2}m_f^2\Big)
+ \frac{3}{2}m_f^4 \ln\Big(\frac{\mu_f+k_f}{m_f}\Big)\Biggr]
\protect\label{eos_P_uqm}
\end{eqnarray}

where,
\begin{eqnarray}
\mu_f=\Big(k_f^2 + m_f^2\Big)^{\frac{1}{2}}
\end{eqnarray}

and the total baryon density is

\begin{eqnarray}
\rho_B=\sum_f \frac{k_f^3}{3\pi^2}
 \protect\label{density_upq}
\end{eqnarray}

 where, $f$ = u, d and s are the quark flavors. The number densities for each flavor are obtained by imposing the charge neutrality condition
\begin{eqnarray} 
 q=\sum_f q_f \rho_f + q_e \rho_e = 0
\end{eqnarray}

and $\beta$ equilibrium condition
\begin{eqnarray}
\mu_d=\mu_u + \mu_e~~~~~~~~ \rm{and}~~~~~~~~~~~ \mu_s=\mu_d
\end{eqnarray}

where the quark chemical potentials are related to the baryon ($\mu_B$) and electron ($\mu_e$) chemical potentials as

\begin{eqnarray} 
\mu_u=(\mu_B - 2\mu_e)/3~~~~~~~~ \rm{and}~~~~~~~~~~~ \mu_s=\mu_d=(\mu_B + \mu_e)/3
\end{eqnarray}

 For SQSs, the Bodmer-Witten conjecture states that the stability is determined in terms of matter energy per baryon $\varepsilon/\rho_B$ which is controlled by the bag pressure \cite{Farhi:1984qu,Torres:2012xv,Ferrer:2015vca}. For SQM to be stable and be the true ground state of the matter, based on Bodmer-Witten conjecture, \cite{Farhi:1984qu,Torres:2012xv,Ferrer:2015vca} estimated the allowed range of $B$ with respect to the stability condition of SQSs demanding that at the surface of the star ($P$=0) the matter energy per baryon $\varepsilon/\rho_B$ of SQM must be less than of the corresponding value for the iron nucleus ($\sim$930 MeV) i.e,

\begin{eqnarray}
\varepsilon/\rho_B \leq 930~ \rm{MeV}
\label{stab_SQS}
\end{eqnarray}

where, $\rho_B$ is the baryon density. In $\beta$ equilibrated SQM, the upper bound on $B$ is set by considering charge neutral 3 flavor SQM while the lower bound is obtained with 2 flavor QM \cite{Torres:2012xv}. Further, model dependent analysis with respect to GW170817 data constrained $B$ with different spin priors for SQSs \cite{Zhou:2017pha}. Refs. \cite{Aziz:2019rgf,Yang:2019rxn} also obtained the allowed range of $B$ for SQSs with a modified Tolman-Oppenheimer-Volkoff (TOV) formalism. However, for HSs the only way to constrain the bag pressure is by model dependent analysis in terms of the various astrophysical constraints \cite{Nandi:2017rhy,Nandi:2020luz}. 
    
\subsection{Density Dependent Bag Pressure}  
\label{DDBP}

As mentioned in the introduction section, at high densities, relevant to compact star cores, the quarks acquire asymptotic freedom and thus to invoke this notion we have considered density dependent bag pressure in the present work following the treatment proposed by \cite{Burgio:2001mk,Burgio:2002sn}. The density dependent bag pressure $B(\rho)$  which attains finite values $B_0~\rm{at}~ \rho=0$ and $B_{as}$ at asymptotic densities is  given by a Gaussian distribution form in terms of $B_0$ and $B_{as}$ as \cite{Burgio:2001mk,Burgio:2002sn}

\begin{eqnarray}
B(\rho) = B_{as} + (B_0 - B_{as})~ \rm{exp}~ [-\beta(\rho/\rho_0)^2]
\label{B}
\end{eqnarray}

where, $\beta$ controls the decrease of $B(\rho)$ from $B_0$ to $B_{as}$ with the increase of density. This form has also been adopted to obtain HS properties by \cite{Miyatsu:2015kwa,Maieron:2004af,Kumar:2023lhv,Pal:2023quk,Sen:2022lig,Sen:2021cgl} and also the stability of SQSs \cite{Isayev:2015jqa,Bordbar:2013uwa,Prasad:2003bw} and proto-QS/HS properties \cite{Bordbar:2020fqj,Bordbar:2020fqj,Yazdizadeh:2013cxa}. We consider the second term of RHS of equation \ref{B} as $\Delta B=B_0 - B_{as}$. 

We intend to show that the variation of each of the three parameters ($B_{as}$, $\Delta B$ and $\beta$) bring substantial changes to the EoS and the structural properties of QSs and HSs in the light of the constraints especially the maximum mass $M_{max}$ and the radius $R_{1.4}$ and tidal deformability $\Lambda_{1.4}$ of a 1.4 $M_{\odot}$ compact star. As we are specifically interested to study the variation of these three parameters that contribute to the density dependence of the bag pressure, we do not include other effects like the perturbative one or repulsive interaction between the quarks or gluon exchange. Moreover, from several analysis, it has been emphasized that the effects of perturbative corrections for the QM interactions on compact star properties can also be realized by changing the bag constant \cite{Glendenning:1997wn}.

\subsection{Hybrid Star Matter}
\label{HSM}

To obtain the EOS of HSM,  we consider the pure hadronic phase to be composed of $\beta$ equilibrated nuclear matter, consisting of the nucleons, electrons and muons, described by the NL3$\omega\rho$6 model \cite{Fortin:2021umb,Grill:2014aea,Pais:2016xiu}. The saturation properties of this model are quite acceptable with respect to various experimental and empirical constraints on them. Theoretical studies suggest that at high density relevant to compact star cores, there may be formation of the hyperons with consequent softening of the EoS and reduction in maximum mass of the star \cite{Glendenning:1997wn}. In the present work, similar to \cite{Liu:2022mje,Bozzola:2019tit,Christian:2018jyd}, we do not consider the formation of hyperons in the hadronic sector due to lack of experimental or observational evidence in support of presence of hyperons in compact star cores. For the pure quark phase, the MIT Bag model is adopted with both density dependent and independent bag pressure as described in the previous section \ref{Quark Phase}.
 
 We assume the surface tension at hadron-quark boundary to be sufficiently large and follow Maxwell construction to invoke phase transition. According to Maxwell criteria, phase transition occurs when the baryon chemical potential and pressure of each of the individual charge neutral phases become equal. With the EoS obtained for both SQSs and HSs for the different values of $B_{as}$, $B_0$ and $\beta$, the structural properties like the gravitational mass $M$ and the radius $R$ of the SQSs and HSs in static conditions are computed by integrating the following Tolman-Oppenheimer-Volkoff (TOV) equations \cite{Tolman:1939jz,Oppenheimer:1939ne}. The dimensionless tidal deformability $\Lambda$ is then calculated following \cite{Hinderer:2007mb,Hinderer:2009ca}.


\section{Result and Discussions}
\label{Results}

\subsection{Quark Stars}
 
We consider the density dependent Bag model following equation \ref{B} since it has already been shown that the original form of Bag model with constant bag pressure cannot satisfy the observational constraints from massive pulsars in case of SQSs unless repulsive or perturbative corrections are included \cite{Lopes:2020btp}.
 
\begin{figure}[!ht]
\centering
\subfloat[]{\includegraphics[width=0.49\textwidth]{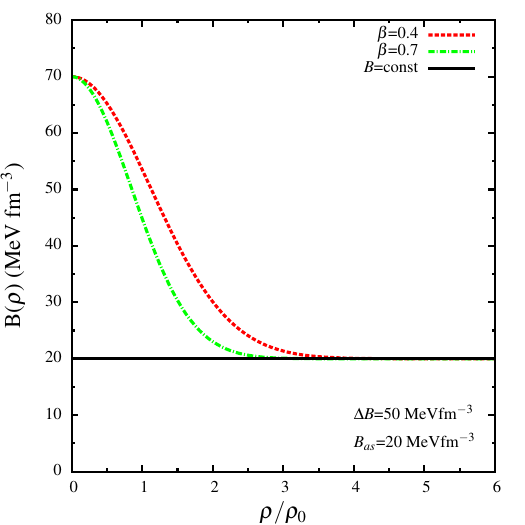}\label{rhoB_beta}}
\hfill
\subfloat[]{\includegraphics[width=0.49\textwidth]{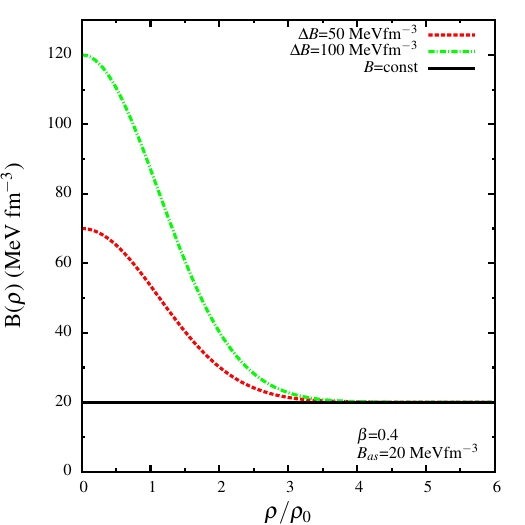}\label{rhoB_delB}}
\caption{(a) Variation of bag pressure with density for different values of $\beta$ fixing $B_{as}$=20 MeV fm$^{-3}$ and $\Delta B$=50 MeV fm$^{-3}$. The density independent (constant) $B$=20 MeV fm$^{-3}$ case is also compared. (b) Variation of bag pressure with density for different values of $\Delta B$ fixing $B_{as}$=20 MeV fm$^{-3}$ and $\beta$=0.4.}
\label{rhoB_beta_delB}
\end{figure}

 We start with concentrating on the density dependent scenario of bag pressure. In figure \ref{rhoB_beta_delB} we show the variation of bag pressure with density, for different values of $\beta$ and $\Delta B$, individually. We find from figure \ref{rhoB_beta} that the asymptotic freedom gained by $B(\rho)$ is earlier in case of $\beta$=0.7 compared to the case when $\beta$=0.4. For $\beta$=0.4, $B(\rho)$ equals $B$=constant around $\rho$=4.12$\rho_0$ while for $\beta$=0.7 the same happens around 3.62$\rho_0$. As we increase the value of $\beta$ further from 0.7, we find that the $M-R$ curves approach more towards that obtained with $B$=constant. Thus in the present work for both SQSs and HSs we choose small values of $\beta<$0.4. Choosing large values of $\Delta B$, we obtain larger deviation from the $B$=constant scenario as seen from figure \ref{rhoB_delB}. Thus in case of both SQSs and HSs, we consider values of $\Delta B \geq$ 50 MeV fm$^{-3}$. Also, lower values of $B_{as}$ shifts the scenario towards the B=constant case. Thus in the present work we choose $B_{as}\geq$ 10 MeV fm$^{-3}$ as the starting value.

\begin{figure}[!ht]
\centering
{\includegraphics[width=0.52\textwidth]{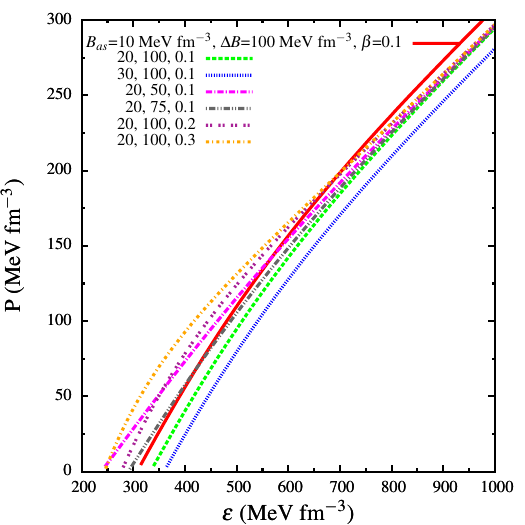}}
\caption{Equation of State of quark stars with density dependent values of bag pressure for variation of ($B_{as}$, $\Delta B$, $\beta$).}
\label{eos_SQS}
\end{figure}

\begin{figure}[!ht]
\centering
\subfloat[]{\includegraphics[width=0.49\textwidth]{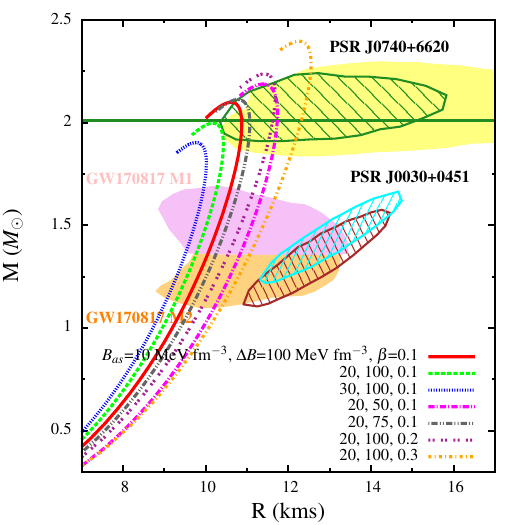}\protect\label{mr_SQS}}
\hfill
\subfloat[]{\includegraphics[width=0.49\textwidth]{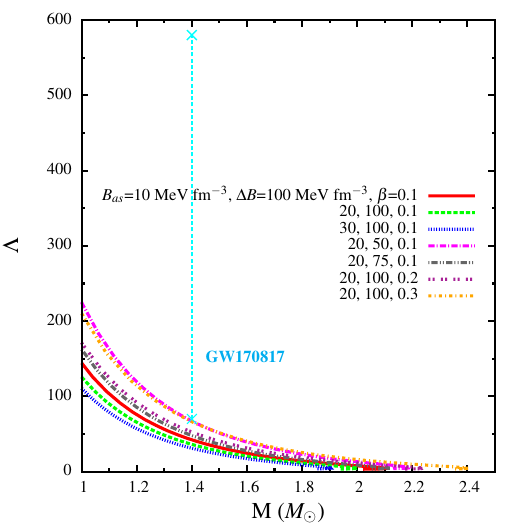}\protect\label{mLam_SQS}}
\caption{(a) Mass-radius relationship quark stars with density dependent values of bag pressure for variation of ($B_{as}$, $\Delta B$, $\beta$). Observational limits imposed from PSR J0740+6620 on maximum mass \cite{Fonseca:2021wxt} and corresponding radius \cite{Miller:2021qha,Riley:2021pdl} are also indicated. The constraints on $M-R$ plane prescribed from GW170817 \cite{LIGOScientific:2018cki}) and NICER experiment for PSR J0030+0451 \cite{Riley:2019yda,Miller:2019cac} are also compared.(b) Variation of tidal deformability with mass for the same. Constraint on $\Lambda_{1.4}$ from GW170817 \cite{LIGOScientific:2018cki} observations is also shown.}
\label{structure_SQS}
\end{figure}

 With the chosen $\beta<$ 0.4, $\Delta B \geq$ 50 MeV fm$^{-3}$ and $B_{as}\geq$ 10 MeV fm$^{-3}$, we checked rigorously the stability conditions of SQSs (equation \ref{stab_SQS}) following the method discussed in section \ref{Quark Phase} and discarded a large number of combinations of ($B_{as}$, $\Delta B$, $\beta$) violating the stability conditions. We varied each parameter at a time - $B_{as}$ in the step of 10 MeV fm$^{-3}$ and $\Delta B$ and $\beta$ in the steps of 50 MeV fm$^{-3}$ and 0.1, respectively. By varying each parameter among ($B_{as}$, $\Delta B$, $\beta$), with such sets that survived the stability test for SQSs, we obtained the EoS and checked rigorously the structural properties of SQSs w.r.t the various recent astrophysical constraints by varying one parameter at a time keeping the other two fixed. To illustrate this, we present in figure \ref{eos_SQS} the EoS and in figure \ref{structure_SQS} the results of the structural properties of SQSs obtained with a few such chosen combinations of ($B_{as}$, $\Delta B$, $\beta$).

 Considering the variation of $B_{as}$ with fixed $(\beta$, $\Delta B)=$ (0.1, 100), the stability condition of SQSs is not satisfied for $B_{as}\geq$ 30 MeV fm$^{-3}$. Within the range (10 $\leq$ $B_{as}\leq$ 30) MeV fm$^{-3}$ and for the chosen fixed values of $\Delta B$ and $\beta$, figure \ref{mr_SQS} shows that for $B_{as}\geq$ 20 MeV fm$^{-3}$ the maximum mass constraint from \cite{Fonseca:2021wxt} is violated. The maximum mass is seen to increase with decreasing values of $B_{as}$ since from figure \ref{eos_SQS} we observe stiffening of the quark EoS for lower values of $B_{as}$ with $\beta$ and $\Delta B$ kept fixed. Next we consider the results for the variation of $\Delta B$ keeping $(B_{as}$, $\beta)=$ (20, 0.1). From figure \ref{mr_SQS} we find that both $M$ and $R$ increase with decreasing values of $\Delta B$ as we notice from figure \ref{eos_SQS} comparative stiffening of the EoS for lower values of $\Delta B$ with fixed $B_{as}$ and $\beta$. Finally we also consider in figure \ref{mr_SQS} the variation of $\beta$ by fixing $(B_{as}$, $\Delta B)=$ (20, 100) where we find that both $M$ and $R$ increase with increasing values of $\beta$. This is also consistent with figure \ref{mr_SQS} which shows that with increasing values of $\beta$ the EoS stiffens for fixed values of $B_{as}$ and $\Delta B$. In figure \ref{mLam_SQS}, we also show the variation of tidal deformability of the SQSs for the variation of each quantity among ($B_{as}$, $\Delta B$, $\beta$) with the other two fixed. Considering figure \ref{structure_SQS} we find that although the constraints on $M-R$ relation from GW170817 \cite{LIGOScientific:2018cki} and PSR J0740+6620 \cite{Fonseca:2021wxt,Miller:2021qha,Riley:2021pdl} are quite satisfied, the NICER data for PSR J0030+0451 and the constraint on $\Lambda_{1.4}$ from GW170817 \cite{LIGOScientific:2018cki} are satisfied with none of the SQS configurations for the different combinations of ($B_{as}$, $\Delta B$, $\beta$).

 Apart from the results presented in figure \ref{structure_SQS}, we have also checked for numerous other combinations of ($B_{as}$, $\Delta B$, $\beta$) and have found that the results for the structural properties of SQS  are more or less  similar to what we have discussed with respect to various astrophysical constraints. The recent astrophysical constraints from NICER data for PSR J0030+0451 and that on $\Lambda_{1.4}$ from GW170817 thus help us to understand that with this form of density dependent bag model, no reasonable QS configuration can be obtained that can satisfy both the stability condition of SQSs and these recent astrophysical constraints simultaneously.
 
We next proceed to study the HS properties with suitable values of ($B_{as}$, $\Delta B$, $\beta$).

\subsection{Hybrid Stars}

We consider the density dependence of bag pressure following equation \ref{B}. Similar to the case of SQSs, we do not consider the values of $B_{as}<$ 10 MeV fm$^{-3}$ since for such lower values of $B_{as}$, $B(\rho)$ approaches more towards the density independent ($B$=constant) case. Due to the same reason we consider $\beta<$ 0.4 and $\Delta B >$ 50 MeV fm$^{-3}$.

\begin{figure}[!ht]
\centering
{\includegraphics[width=0.52\textwidth]{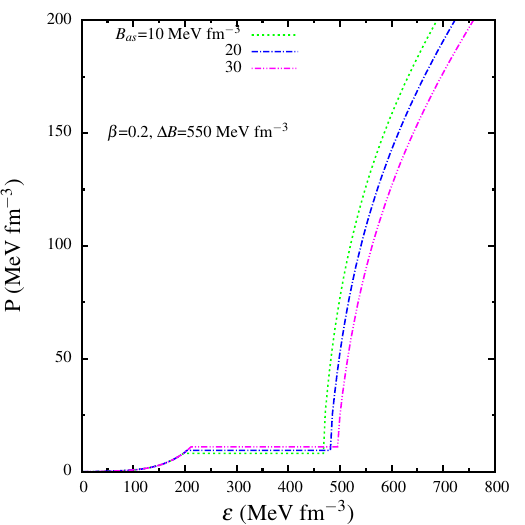}}
\caption{Equation of State of hybrid stars with density dependent values of bag pressure for variation of $B_{as}$ keeping $\Delta B$=550 MeV fm$^{-3}$ and $\beta=$0.2.}
\label{eos_HS_Bas}
\end{figure} 

\begin{figure}
\centering
\subfloat[]{\includegraphics[width=0.49\textwidth]{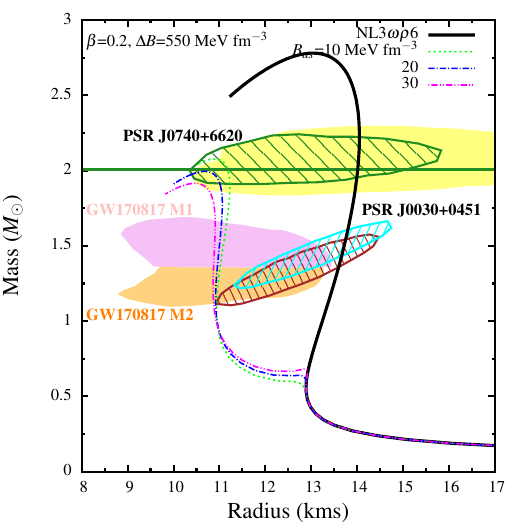}\protect\label{hyb_mr_Bas}}
\hfill
\subfloat[]{\includegraphics[width=0.49\textwidth]{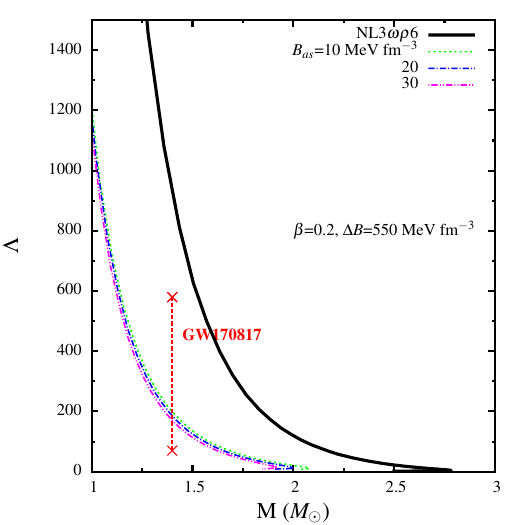}\protect\label{hyb_mLam_Bas}}
\caption{(a) Mass-radius relationship of hybrid stars with density dependent values of bag pressure for variation of $B_{as}$ keeping $\Delta B$=550 MeV fm$^{-3}$ and $\beta=$0.2. (b) Variation of tidal deformability with mass for the same.}
\label{hyb_mrlam_Bas}
\end{figure}
 
 We first try to check the possible values of $B_{as}$ suitable for obtaining a reasonable HS configuration in the light of the various astrophysical constraints. In figure \ref{eos_HS_Bas} we display the EoS of HSs for variation of $B_{as}$ keeping $\Delta B$=550 MeV fm$^{-3}$ and $\beta=$0.2. We find that transition density is very less affected for the variation of $B_{as}$. However, it is slightly higher for higher values of $B_{as}$. The same is reflected in the corresponding mass-radius relationship in figure \ref{hyb_mr_Bas}. In figure \ref{hyb_mrlam_Bas} we show the dependence of the structural properties of HSs on $B_{as}$ for $(\Delta B$, $\beta)=$ (550, 0.2). From figure \ref{hyb_mr_Bas} we find that for $(\Delta B$, $\beta)=$ (550, 0.2), $M_{max}$ increases while the transition density and the transition mass $M_t$ decreases with decreasing values of $B_{as}$. From figure \ref{eos_HS_Bas} we notice that although the transition density is not much affected by $B_{as}$, soon after the transition the EoS shows considerable stiffening with decreasing values of $B_{as}$. Since in this case the maximum mass of the HSs lie in the second (hybrid) branch, $M_{max}$ also increases with decreasing values of $B_{as}$. However, the only value of $B_{as}$ that satisfies all the constraint on the $M-R$ plane is $B_{as}=$ 10 MeV fm$^{-3}$. Both $B_{as}=$ 20 and 30 MeV fm$^{-3}$ do not satisfy the maximum mass constraint despite satisfying the other constraints on the $M-R$ relation. Thus we see that the suitable values of $B_{as}$ is also quite limited to a small range around 10 MeV fm$^{-3}$. From figure \ref{hyb_mLam_Bas}, it is clear that for all the chosen values of $B_{as}$, the HS configurations satisfy the constraint on $\Lambda_{1.4}$ from GW170817.
 
\begin{figure}[!ht]
\centering
{\includegraphics[width=0.52\textwidth]{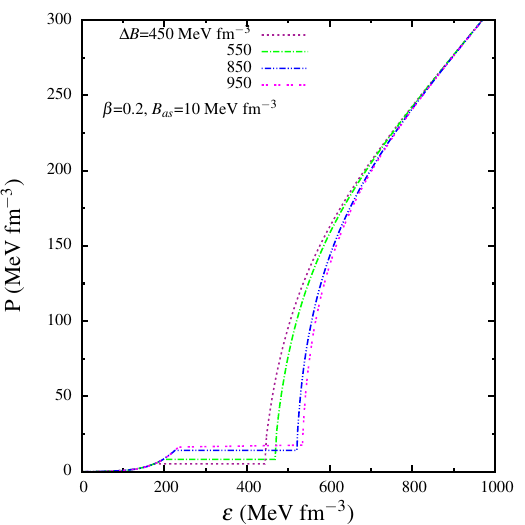}}
\caption{Equation of State of hybrid stars with density dependent values of bag pressure for variation of $\Delta B$ keeping $B_{as}$=10 MeV fm$^{-3}$ and $\beta=$0.2.}
\label{eos_HS_delB}
\end{figure}

\begin{figure}
\centering
\subfloat[]{\includegraphics[width=0.49\textwidth]{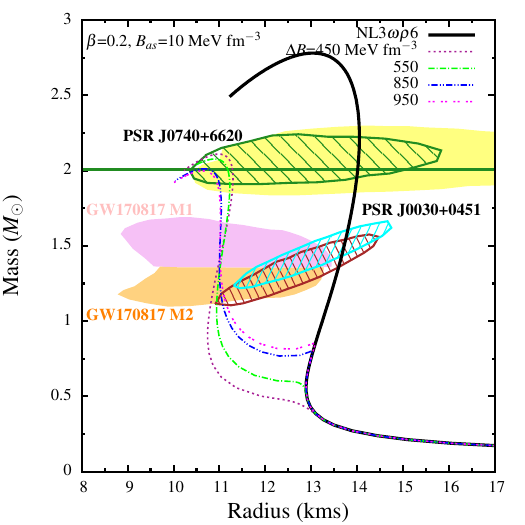}\protect\label{Hybrid_mr_delB}}
\hfill
\subfloat[]{\includegraphics[width=0.49\textwidth]{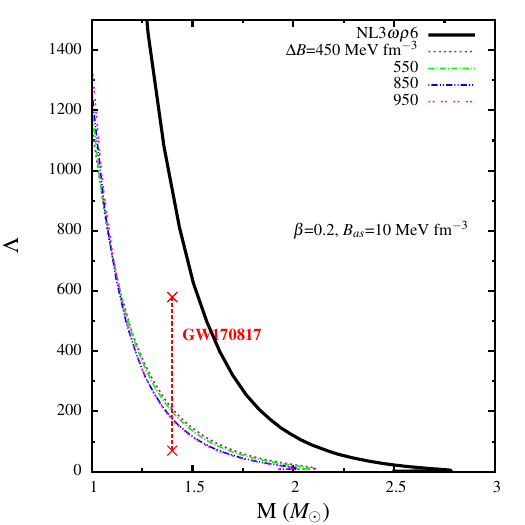}\protect\label{Hybrid_mLam_delB}}
\caption{(a) Mass-radius relationship of hybrid stars with density dependent values of bag pressure for variation of $\Delta B$ keeping $B_{as}$=10 MeV fm$^{-3}$ and $\beta=$0.2. (b) Variation of tidal deformability with mass for the same.}
\label{Hybrid_mrLam_delB}
\end{figure}

With $(B_{as}$, $\beta)=$ (10, 0.2), we next proceed to obtain suitable values of $\Delta B$ with which the HS configurations can satisfy the various astrophysical constraints. The corresponding EoS of the HS is shown in figure \ref{eos_HS_delB}. Compared to the variation of $(B_{as}$ in figure \ref{eos_HS_Bas}, we find from figure \ref{eos_HS_delB} $\Delta B$ has noticeable effect on the transition density which decreases with decreasing values of $\Delta B$. The quark phase followed by the phase transition region shows an interesting feature. The EoS in this phase stiffens with decreasing values of $\Delta B$ upto a certain density after which the EoS for different $\Delta B$ merge, indicating the asymptotic density. This is also consistent with figure \ref{rhoB_delB} where  the bag pressure $B(\rho)$ for two different values of $\Delta B$ is seen to merge at the asymptotic density. From figure \ref{Hybrid_mr_delB}, we find that for $(B_{as}$, $\beta)=$ (10, 0.2), $M_{max}$ increases while the transition density and $M_t$ decreases with decreasing values of $\Delta B$. It is clear that all the constraints on the $M-R$ plane is satisfied when $\Delta B=$ (550 - 850) MeV fm$^{-3}$. Below $\Delta B=$ 550 MeV fm$^{-3}$ the NICER data is violated while above 850 MeV fm$^{-3}$ the lower bound on maximum mass is not fulfilled. Thus unlike $B_{as}$ and $\beta $, we find a reasonable range of $\Delta B$ for which the HS configurations satisfy the present day astrophysical constraints. Since in the present work we introduce the effects of $B_0$ via $\Delta B$, unlike \cite{Burgio:2002sn} we find that this parameter is not only important for determining the structural properties of HSs but also the transition density. For $(B_{as}$, $\beta)=$ (10, 0.2), hadron-quark crossover and thus phase transition is not obtained for $\Delta B\leq$ 250 MeV fm$^{-3}$. In figure \ref{Hybrid_mLam_delB}, we see that the constraint on $\Lambda_{1.4}$ from GW170817 is satisfied for all the HS configurations for different values of $\Delta B$ by fixing $(B_{as}$, $\beta)=$ (10, 0.2). This constraint is satisfied even for $\Delta B$=450 and 950 MeV fm$^{-3}$. 
 
\begin{figure}[!ht]
\centering
{\includegraphics[width=0.52\textwidth]{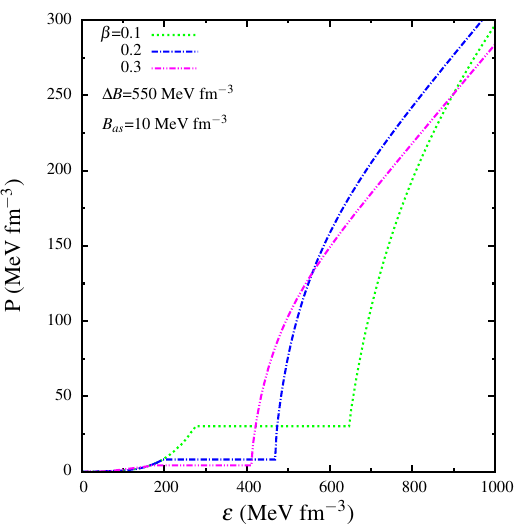}}
\caption{Equation of State of hybrid stars with density dependent values of bag pressure of $\beta$ keeping $B_{as}$=10 MeV fm$^{-3}$ and $\Delta B$=550 MeV fm$^{-3}$.}
\label{eos_HS_beta}
\end{figure}

\begin{figure}
\centering
\subfloat[]{\includegraphics[width=0.49\textwidth]{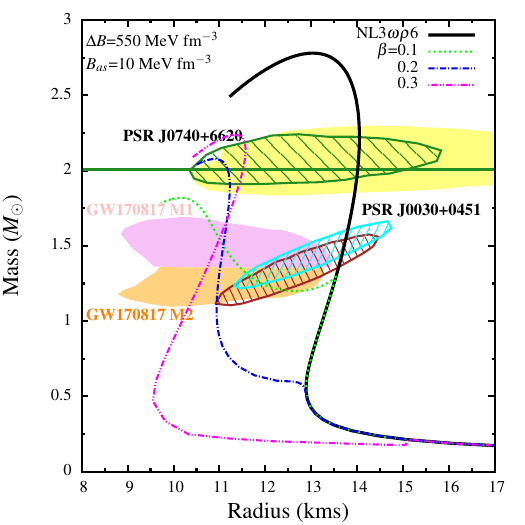}\protect\label{Hybrid_mr_beta}}
\hfill
\subfloat[]{\includegraphics[width=0.49\textwidth]{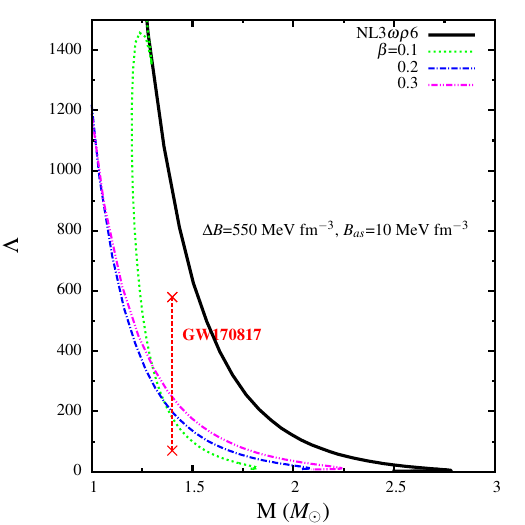}\protect\label{Hybrid_mLam_beta}}
\caption{(a) Mass-radius relationship of hybrid stars with density dependent values of bag pressure for variation of $\beta$ keeping $B_{as}$=10 MeV fm$^{-3}$ and $\Delta B$=550 MeV fm$^{-3}$. (b) Variation of tidal deformability with mass for the same.}
\label{Hybrid_mrLam_beta}
\end{figure}

 We finally try to constrain the value of $\beta$ with HS configurations keeping $(B_{as}$, $\Delta B)=$ (10, 550) fixed as shown in figures \ref{eos_HS_beta} and \ref{Hybrid_mrLam_beta}. From the hybrid EoS shown in figure \ref{eos_HS_beta} and the corresponding mass-radius variation shown in figure \ref{Hybrid_mr_beta}, we notice that the maximum mass increases while the transition density and $M_t$ decreases with increasing value of $\beta$. It is clear that the only value of $\beta$ that satisfies all the constraint on the $M-R$ plane is 0.2. For $\beta$=0.1 the HS configuration does not satisfy the maximum mass constraint despite satisfying the GW170817 and the NICER data while for $\beta$=0.3 the NICER data is violated. So in case of HSs, the allowed value of $\beta$ is quite restricted within a narrow window in the vicinity of 0.2. In figure \ref{Hybrid_mLam_beta}, we find that the constraint on $\Lambda_{1.4}$ from GW170817 is satisfied for all the HS configurations for different values of $\beta$ keeping $(B_{as}$, $\Delta B)$= (10, 550).

 Interestingly, from figures \ref{hyb_mr_Bas} and \ref{Hybrid_mr_delB}, we note that in cases of variation of $B_{as}$ and $\Delta B$, we obtain twin star configurations with the existence of two maximas - one on the hadronic phase branch and the other on the quark phase branch as seen in \cite{Pal:2023quk,Sen:2022lig,Laskos-Patkos:2023cts,Tsaloukidis:2022rus,Gorda:2022lsk}. A few configurations also show a region of instability in between the two phases following phase transition. Such region is seen when the transition is at slightly higher density and corresponds to the points when $dM/d\varepsilon_c <0$, where $\varepsilon_c$ is the central energy density. Unlike the case of variation of $\beta$ in figures \ref{Hybrid_mr_beta}, existence of special points is noted on the $M-R$ relation of HSs in case of variation of $B_{as}$ in figure \ref{hyb_mr_Bas} and $\Delta B$ in figures \ref{Hybrid_mr_delB}. A special point on the $M-R$ plot of HSs indicates a small region where all the HS solutions merge irrespective of the different transition densities for different values of bag pressure. This feature is also noted in \cite{Cierniak:2020eyh,Yudin:2014mla,Pal:2023quk,Sen:2022lig} in the context of formation of hybrid and twin stars.
 
 Overall, we find that the three parameters affect the properties of both SQSs and HSs. For the increasing values of the parameter $B_{as}$, that carries the notion of asymptotic freedom of the quarks at particular high density, $M_{max}$ decreases for both SQSs and HSs. Similar effect is noticed for the parameter $\Delta B$ that bears the essence of the bag pressure at vanishing density ($B_0$). The opposite trend is noticed for the parameter $\beta$. This parameter regulates the decrease of $B(\rho)$ from $B_0$ to $B_{as}$ as seen from figure \ref{rhoB_beta}. With fixed values of $B_{as}$ and $\Delta B$, the quarks acquire asymptotic freedom comparatively early for a higher value of $\beta$. Therefore it can be said that we obtain massive SQS and HS configurations when the quarks gain early asymptotic freedom through a higher value of $\beta$.


\section{Summary and Conclusion}
\label{Conclusion}

 We analyzed of the structural properties of SQSs in the framework of MIT Bag model by considering the bag pressure to be density dependent $B(\rho)$. The density dependence of the bag pressure is obtained by using a Gaussian distribution form involving the parameters $B_{as}$, $\Delta B$ and $\beta$. We checked the stability conditions of SQSs rigorously for various combinations of ($B_{as}$, $\Delta B$, $\beta$) and those which survived the stability test could not satisfy all the astrophysical constraints. Hence we conclude that within the framework of this form of density dependent bag model, reasonable SQS configurations cannot be obtained that can simultaneously satisfy the stability condition (Bodmer-Witten conjecture) and the recent astrophysical constraints from GW170817 and NICER data for PSR J0030+0451. We found that mostly the NICER data for PSR J0030+0451 and the one on $\Lambda_{1.4}$ from GW170817 serve as excellent tools to constrain the parameters of the density dependent bag model. 
  
 We also extended our work to obtain the structural properties of HSs. In the $B$=constant scenario,  we obtained no suitable value of $B$ for which the HSs could satisfy all the present day astrophysical constraints.  For the variation of $B_{as}$ and $\Delta B$, we notice distinct special points on the $M-R$ relations for different values of bag pressure. Considering the different astrophysical constraints, they are satisfied by the HS configurations for The combinations  of $B_{as}$, $\Delta B$ and $\beta$ whoch satisfied different astrophysical constraints are as follows 
 
\begin{itemize}
\item For $B_{as}$=10 MeV fm$^{-3}$ and $\Delta B$=550 MeV fm$^{-3}$, $\beta$ is around 0.2 but not 0.1 or less or 0.3 or more.
\item For $\Delta B$=550 MeV fm$^{-3}$ and $\beta$=0.2, $B_{as}$ is around 10 MeV fm$^{-3}$ but not 20 MeV fm$^{-3}$ or more.
\item For $B_{as}$=10 MeV fm$^{-3}$ and $\beta$=0.2, $\Delta B$=(550 - 850) MeV fm$^{-3}$.
\end{itemize} 

 From the above results obtained for HSs with the NL3$\omega\rho$6 model for the hadronic phase, we conclude that similar to the case of SQSs we find that the maximum mass increases with increasing value of $\beta$ while the opposite behavior is seen in case of both $B_{as}$ and $\Delta B$. Unlike the case of SQS configurations, (obtained with density dependent bag pressure) which do not satisfy all the astrophysical constraints, the HSs fulfill such constraints within a very restricted range of $B_{as}$ and $\beta$ and a little more flexible range of $\Delta B$.  Also, unlike the case of SQSs, the value of $\Lambda_{1.4}$ do not put any strong constraint on the choice of the 3 parameters in case of HSs. The restrictions on the later case are mainly obtained from the constraints on $M_{max}$, GW170817 data for the $M-R$ values and the NICER data for PSR J0030+0451. 
 
 Thus we broadly conclude that the present form of density dependent bag model is suitable for obtaining reasonable hybrid star (HS) configurations within some restricted range of the parameters but not strange quark stars (SQS) in view of the recent astrophysical constraints on the structural properties of compact stars.
 
\section*{Acknowledgements}
Work of DS was supported by the NRF research Grants (No. 2018R1A5A1025563).


\bibliographystyle{spphys}

\end{document}